\documentclass[superscriptaddress,aps,prd,nofootinbib,twocolumn,showpacs]{revtex4-1}

\usepackage{multirow}
\usepackage{amsmath,epsfig,ulem}
\usepackage{amsfonts}
\usepackage{amsmath}
\usepackage{amssymb}
\usepackage{graphicx}
\usepackage{hhline,color}
\usepackage{pifont}
\usepackage{lmodern}
\usepackage[utf8]{inputenc}

\newcommand{\unit}[1]{\ensuremath{\, \mathrm{#1}}}

\newcommand{\pc}[1]{\ensuremath{\left(#1\right)}}

\newcommand{\ev}[1]{\ensuremath{\left\langle #1\right\rangle}}

\def\beq{\begin{equation}}
\def\eeq{\end{equation}}
\def\beqa{\begin{eqnarray}}
\def\eeqa{\end{eqnarray}}
\def\ban{\begin{eqnarray*}}
\def\ean{\end{eqnarray*}}
\def\bi{\begin{itemize}}
\def\ei{\end{itemize}}

\newcommand{\Z}{\mathbb{Z}}

\begin{document}

\title{Strange quark chiral phase transition in hot 2$+$1-flavor magnetized quark matter}

\author{M\'arcio Ferreira}
\email{mferreira@teor.fis.uc.pt}
\affiliation{Centro de F\'{\i}sica Computacional, Department of Physics,
University of Coimbra, P-3004$-$516  Coimbra, Portugal}
\author{Pedro Costa}
\email{pcosta@teor.fis.uc.pt}
\affiliation{Centro de F\'{\i}sica Computacional, Department of Physics,
University of Coimbra, P-3004$-$516  Coimbra, Portugal}
\author{Constan\c ca Provid\^encia}
\email{cp@teor.fis.uc.pt}
\affiliation{Centro de F\'{\i}sica Computacional, Department of Physics,
University of Coimbra, P-3004$-$516  Coimbra, Portugal}

\date{\today}

\begin{abstract}
Using the Polyakov--Nambu--Jona-Lasinio model the strange quark chiral phase transition and the
effect of its current mass on a hot magnetized three flavor quark matter at zero chemical 
potential is investigated. 
The impact of the 't Hooft mixing term on the restoration of the chiral symmetry on the light 
and strange sectors is studied and the critical temperature dependence on the magnetic field
strength of the chiral strange transition is analyzed.
It is shown that the $s$ quark is much less sensitive to the magnetic field than the light quarks. 
Due to the 't Hooft term, it has a strong influence on the light quarks at all temperatures for 
small magnetic fields and for temperatures close to the transition temperature at zero magnetic field
and above for strong magnetic fields. In particular, the large mass of the $s$ quark makes the 
chiral transition of the light sector smoother and shifted to larger temperatures.
A scalar coupling that weakens when the magnetic field increases will originate an inverse magnetic 
catalysis also in the $s$ quark and smoothen the signature of the crossover on thermodynamical 
quantities such as the sound velocities or the specific heat.

\end{abstract}

\pacs{24.10.Jv, 11.10.-z, 25.75.Nq}

\maketitle

\section{Introduction}

Strangeness is a very important degree of freedom that must be considered 
when discussing the QCD phase diagram.
The up, down and strange quark masses control the amount of explicit chiral 
symmetry breaking in QCD. However, in the real world, the strange quark is 
significantly heavier than the nonstrange quarks which makes the physics 
related with the strange quark very interesting once the SU(3)-flavor symmetry 
is  explicitly broken by its mass. 
In fact,  chiral symmetry is a very important concept in the up/down sector 
and even, although with larger deviations, when strange quarks are also included 
\cite{Buballa:2003qv}: when two massless quarks are considered (and the strange quark 
mass is taken to be infinite) the chiral phase transition is of second order, but, in 
a world with three massless quarks, the chiral phase transition is of first order. 
If the strange quark mass is reduced from infinity to zero, at some point the 
phase transition must change from second order to first order and there must be 
a tricritical strange quark mass, $m^{tric}_s$, where the second order chiral 
transition ends and the first order region begins \cite{mscrit}.

The relevance of strangeness is transversal to all regions across the phase diagram.
In the interior of a neutron star (high density and low temperature region) it is expected 
that strangeness is present either in the form of hyperons, a kaon condensate or a 
core of deconfined quark matter \cite{glendenning2000}. 
The recent measurement of the mass of the two solar mass  millisecond pulsars 
PSR J1614$-$2230 \cite{demorest2010} and  PSR J1903$+$0327 \cite{antoniadis2012} 
places quite strong constraints on the core composition of neutron stars. 
The compatibility of these large masses with the appearance of strangeness has been 
questioned on the basis of microscopic approaches to the hadronic equation of state 
\cite{bhf,vidana2011}.  
Within a relativistic mean field approach it has been shown that it is still possible to
accommodate these large masses even considering the presence of hyperons or kaons 
(see for instance \cite{cavagnoli2011,bednarek2012,weissenborn2012,providencia2013,panda2014}), 
since there is a large uncertainty on the coupling of hyperons to nucleons. 
Another possibility is that the interior of the neutron star contains a quark core 
\cite{bonanno2011}.

In relativistic heavy-ion collisions the strange and multistrange particle production 
is an important tool to investigate the properties of the hot and dense matter created in 
the collision, since there is no net strangeness content in the initially colliding nuclei 
\cite{Elia:2013fma}.
An enhanced production of strange particles in A$-$A compared to $pp$ 
collisions  was one of the first signatures proposed for the deconfined quark-gluon plasma  
\cite{Rafelski:1982pu,Koch:1986ud}.
Very recently,  the possibility of multiple chemical freeze-outs was suggested, in particular,  
the strange freeze-out which would indicate a clear separation of pion and kaon chemical freeze-outs 
\cite{Chatterjee:2013yga}. Based on known systematics of hadron cross sections, it was argued that 
different particles can freeze out of the fireball produced in heavy-ion collisions at different 
times. Another alternative approach to treat the strange particle freeze-out separately, with the 
full chemical equilibrium, was presented in \cite{Bugaev:2013sfa}: based on the conservation laws, 
the connection between the freeze-outs of strange and nonstrange hadrons was achieved.
Strangeness freeze-out in heavy-ion collisions is also deserving the attention of lattice QCD (LQCD)
community. 
It was found that experimentally unobserved strange hadrons become thermodynamically
relevant in the vicinity of the QCD crossover, modifying the yields of the ground state strange hadrons 
in heavy-ion collisions, which leads to significant reductions in the chemical freeze-out temperature 
of strange hadrons \cite{Bazavov:2014xya}.  However, the question whether of hadrons of different quark 
composition freeze out simultaneously or exhibit a flavor hierarchy \cite{Bellwied:2013cta} still has 
no answer.

Another relevant aspect of strangeness in the phase diagram refers to the determination of  
the confinement/deconfinement pseudocritical temperature.
At finite temperature and zero chemical potential, LQCD results indicate a crossover from the 
hadronic phase to the quark-gluon plasma for realistic $u$, $d$ and $s$ quark masses \cite{Aoki:2006we}
and there are different prescriptions which lead to different pseudocritical temperatures for both, 
the chiral and the confinement/deconfinement phase transitions. 
For the deconfinement transition a way to define the pseudocritical point is to use the peak position 
of the Polyakov loop susceptibility. 
However, instead of the Polyakov loop, it is also possible to use the strange quark number 
susceptibility, 
$\chi_s=\frac{T}{V}\frac{\partial^2(\mbox{ln} Z)}{\partial\mu^2_s}$,
to define the pseudocritical temperature, being $\mu_s$ the chemical potential for strange quarks.
As pointed out in \cite{Fukushima:2013rx}, $\chi_s$  behaves in a similar way to 
the Polyakov loop: in the Nambu--Jona-Lasinio (NJL) model coupled to the Polyakov loop (PNJL), 
when the quark mass of the heavy flavor is large enough, the susceptibility $\chi_s$ is proportional 
to the Polyakov loop, which makes this quantity qualified as an order parameter \cite{Fukushima:2013rx}. 
So, the inflection point of $\chi_s$ gives the pseudocritical temperature consistent with the use of the 
peak position of the Polyakov loop susceptibility.
Lattice data from different collaborations show that the two pseudocritical temperatures are close 
to each other \cite{Borsanyi:2010bp,Bazavov:2011nk}, making this behavior a general trend.
In the framework of lattice QCD calculations, the strange quark number susceptibility is also a very interesting quantity 
from the theoretical point of view because it is related to a conserved current, thus no renormalization 
ambiguities appear, which makes direct comparisons particularly easy \cite{Borsanyi:2010bp}. 

Recent LQCD results also suggest that the deconfinement of strangeness takes place at the chiral crossover region,
and for temperatures larger than twice the chiral crossover temperature, the strangeness carrying degrees of
freedom inside the quark-gluon plasma can be described by a weakly interacting gas of quarks 
\cite{Bazavov:2013dta}.
Finally, the direct determination of the light and strange quark condensates from full LQCD was performed 
in \cite{McNeile:2012xh}.

In heavy-ion collisions it is also important to consider the presence of magnetic fields.
Although time dependent and short lived \cite{Kharzeev}, the magnetic fields involved can reach
intensities of the order $eB = 5 - 30$ $m_{\pi}^2$ (corresponding to $1.7 \times 10^{19} - 10^{20}$ gauss)
and temperatures varying from $T = 120-200$ MeV.
For example, the estimated value of the magnetic field strength for the LHC energy 
is of the order $eB \sim 15$ $m^2_{\pi}$ \cite{Skokov:2009qp}.

At zero chemical potential and finite temperature, when the effect of an external magnetic field in 
QCD matter with $N_f=2+1$ flavors with physical quark masses is taken into account, 
LQCD results show that light and heavy quark sectors respond differently to the magnetic field
\cite{Bali:2011qj}.
The magnetic field suppresses the light quark condensates near the transition temperature,
giving them a nonmonotonic behavior with $eB$. However, according to \cite{Bali:2012zg}, 
the $s$ quark condensate increases with $eB$ for all temperatures.
Furthermore, it was observed in \cite{Bali:2011qj} that the pseudocritical 
temperatures for $N_f=3$ heavy flavors do not change much with the magnetic field.
The Polyakov loop also reacts to the magnetic field: it increases sharply with the 
magnetic field around the transition temperature and the transition temperature
taken from the renormalized Polyakov loop clearly decreases with the magnetic field
\cite{Bruckmann:2013oba}.

Low-energy effective models, namely the NJL and PNJL in 2+1 flavors, have also been used to study 
the influence of an external magnetic field in the QCD phase diagram at zero chemical potential 
and finite temperature \cite{Fu:2013ica,Ferreira:2013tba,Ferreira:2013oda}. 
At finite temperature and chemical potential the combined effects of the strangeness, isospin 
asymmetry and an external magnetic field on the location of the critical end point (CEP) in the 
QCD phase diagram were investigated \cite{Costa:2013zca}.
It was shown that isospin asymmetry shifts the CEP to larger baryonic
chemical potentials and smaller temperatures, and that  at large
asymmetries the CEP disappears. However, a strong enough magnetic
field drives the system into a first order phase transition. It was
also discussed that strangeness shifts the CEP to larger baryonic
densities and in most cases also to larger baryonic chemical potentials.

Almost all low-energy effective models, at zero chemical potential, including the 
NJL-type models, find an enhancement of the condensate due 
to the magnetic field, the so-called magnetic catalysis, and no reduction of 
the pseudocritical chiral transition temperature with the magnetic field 
\cite{Gatto:2012sp,Fraga:2012rr,D'Elia:2012tr}.
However, recent studies using the NJL and PNJL \cite{Ferreira:2014kpa} could reproduce 
the inverse magnetic catalysis (IMC) effect predicted by LQCD, if a magnetic field 
dependent scalar coupling that decreases when the field increases is considered.
This dependence of the coupling allows us to reproduce the LQCD results with respect to 
the quark condensates and to the Polyakov loop: due to the magnetic field the quark 
condensates are enhanced at low and high temperatures and suppressed for temperatures 
close to the transition temperature \cite{Bali:2011qj,Bali:2012zg}.

In this paper we will investigate the strange quark chiral phase transition 
in a hot 2$+$1-flavor magnetized quark matter and  identify the features of the 
QCD phase diagram due to the presence of strangeness.
The main property that distinguishes the $u$ and $d$ quarks from the $s$ quark 
is its mass, more than one order of magnitude larger.
Moreover, a term like the 't Hooft term, that mixes flavors, will  have a 
significant effect, and the behavior of the $u$ and $d$ quarks will be strongly 
influenced by the $s$ quark.
We will analyze how these features affect the QCD phase diagram and will identify 
the importance of including the strangeness degree of freedom.

\section{Model and Formalism}
\label{sec:model}

\subsection{Model Lagrangian and gap equations}

We describe three flavor ($N_c=3$) quark matter subject to strong magnetic fields 
within the 2+1 PNJL model.
The PNJL Lagrangian with explicit chiral symmetry breaking, where the quarks couple 
to a (spatially constant) temporal background gauge field, represented in terms of the
Polyakov loop, and in the presence of an external magnetic field is given by 
\cite{Fukushima:2003fw.Ratti:2005jh}:
\begin{eqnarray}
{\cal L} &=& {\bar{q}} \left[i\gamma_\mu D^{\mu}-
	{\hat m}_f \right ] q ~+~ {\cal L}_{sym}~+~{\cal L}_{det} \nonumber\\
&+& \mathcal{U}\left(\Phi,\bar\Phi;T\right) - \frac{1}{4}F_{\mu \nu}F^{\mu \nu},
	\label{Pnjl}
\end{eqnarray}
where the quark sector is described by the  SU(3) version of the NJL model which
includes scalar-pseudoscalar and the 't Hooft six fermion interactions, that
models the axial $U_A(1)$ symmetry breaking  \cite{Hatsuda:1994pi.Klevansky:1992qe},
with ${\cal L}_{sym}$ and ${\cal L}_{det}$  given by \cite{Buballa:2003qv}
\begin{eqnarray}
	{\cal L}_{sym}= G_s \sum_{a=0}^8 \left [({\bar q} \lambda_ a q)^2 + 
	({\bar q} i\gamma_5 \lambda_a q)^2 \right ] ,
\end{eqnarray}
\begin{eqnarray}
	{\cal L}_{det}=-K\left\{{\rm det} \left [{\bar q}(1+\gamma_5)q \right] + 
	{\rm det}\left [{\bar q}(1-\gamma_5)q\right] \right \}
\end{eqnarray}
where $q = (u,d,s)^T$ represents a quark field with three flavors, 
${\hat m}_f= {\rm diag}_f (m_u,m_d,m_s)$ is the corresponding (current) mass matrix,
$\lambda_0=\sqrt{2/3}I$  where $I$ is the unit matrix in the three flavor space
and $0<\lambda_a\le 8$ denote the Gell-Mann matrices.
The coupling between the (electro)magnetic field $B$ and quarks, and between the 
effective gluon field and quarks, is implemented  {\it via} the covariant derivative 
$D^{\mu}=\partial^\mu - i q_f A_{EM}^{\mu}-i A^\mu$ where $q_f$ represents the 
quark electric charge ($q_d = q_s = -q_u/2 = -e/3$),  $A^{EM}_\mu$ and 
$F_{\mu \nu }=\partial_{\mu }A^{EM}_{\nu }-\partial _{\nu }A^{EM}_{\mu }$ 
are used to account for the external magnetic field and 
$A^\mu(x) = g_{strong} {\cal A}^\mu_a(x)\frac{\lambda_a}{2}$ where
${\cal A}^\mu_a$ is the SU$_c(3)$ gauge field.
We consider a  static and constant magnetic field in the $z$ direction, 
$A^{EM}_\mu=\delta_{\mu 2} x_1 B$.
In the Polyakov gauge and at finite temperature the spatial components of the 
gluon field are neglected: 
$A^\mu = \delta^{\mu}_{0}A^0 = - i \delta^{\mu}_{4}A^4$. 
The trace of the Polyakov line defined by
$ \Phi = \frac 1 {N_c} {\langle\langle \mathcal{P}\exp i\int_{0}^{\beta}d\tau\,
A_4\left(\vec{x},\tau\right)\ \rangle\rangle}_\beta$
is the Polyakov loop which is the {\it exact} order parameter of the $\Z_3$ 
symmetric/broken phase transition in pure gauge.

The coupling constant $G_s$ in ${\cal L}_{sym}$ denotes the scalar-type four-quark interaction 
of the NJL sector. Since the model is not renormalizable, we use as a regularization scheme
a sharp cutoff in three-momentum space,  $\Lambda$, only for the divergent ultraviolet integrals
(the details can be found in Ref. \cite{Menezes:2008qt.Menezes:2009uc}). 
The parameters of the model, $\Lambda$, the coupling constants $G_s$ and $K$
and the current quark masses $m_u$, $m_d$ and $m_s$ are determined  by fitting
$f_\pi$, $m_\pi$ , $m_K$, and $m_{\eta'}$ to their empirical values. 
We consider $\Lambda = 602.3 \, {\rm MeV}$ , $m_u= m_d=\,  5.5 \,{\rm MeV}$,
$m_s=\,  140.7\, {\rm MeV}$, $G \Lambda^2= 1.385$ and $K \Lambda^5=12.36$
as in \cite{Rehberg:1995kh}.

To describe the pure gauge sector an effective potential $\mathcal{U}\left(\Phi,\bar\Phi;T\right)$
is chosen in order to reproduce the results obtained in lattice calculations \cite{Roessner:2006xn}:
\begin{eqnarray}
	& &\frac{\mathcal{U}\left(\Phi,\bar\Phi;T\right)}{T^4}
	= -\frac{a\left(T\right)}{2}\bar\Phi \Phi \nonumber\\
	& &
	+\, b(T)\mbox{ln}\left[1-6\bar\Phi \Phi+4(\bar\Phi^3+ \Phi^3)-3(\bar\Phi \Phi)^2\right],
	\label{Ueff}
\end{eqnarray}
where $a\left(T\right)=a_0+a_1\left(\frac{T_0}{T}\right)+a_2\left(\frac{T_0}{T}\right)^2$, 
$b(T)=b_3\left(\frac{T_0}{T}\right)^3$.
The standard choice of the parameters for the effective potential $\mathcal{U}$ is
$a_0 = 3.51$, $a_1 = -2.47$, $a_2 = 15.2$, and $b_3 = -1.75$.

As is well known, the effective potential exhibits the feature of a phase transition from 
color confinement ($T<T_0$, the minimum of the effective potential being at $\Phi=0$) to color
deconfinement ($T>T_0$, the minimum of the effective potential occurring at $\Phi \neq 0$).

We know that the parameter $T_0$ of the Polyakov potential defines the onset of deconfinement
and is normally fixed to $270$ MeV according to the critical temperature for the deconfinement
in pure gauge lattice findings (in the absence of dynamical fermions) \cite{Kaczmarek:2002mc}. 
When quarks are added to the system, quark backreactions must be taken into account,
thus a decrease in $T_0$ to $210\unit{MeV}$ is required to obtain the deconfinement
pseudocritical temperature given by LQCD, within the PNJL model. 
Therefore, the value of $T_0$ is fixed in order to reproduce LQCD results 
($\sim$ 170 MeV \cite{Aoki:2009sc}).

The thermodynamical potential for the three flavor quark sector $\Omega$ is written as
\begin{align}
\Omega(T,\mu)&=G_s \sum_{i=u,d,s}\ev{\bar{q}_iq_i}^2
+4K\ev{\bar{q}_uq_u}\ev{\bar{q}_dq_d}\ev{\bar{q}_sq_s} \nonumber \\
+&{\cal U}(\Phi,\bar{\Phi},T)+\sum_{i=u,d,s}\pc{\Omega_{\text{vac}}^i+\Omega_{\text{med}}^i
+\Omega_{\text{mag}}^i}
\end{align}
with the flavor contributions from vacuum $\Omega_{\text{vac}}^i$, medium $\Omega_{\text{med}}^i$ 
and magnetic field $\Omega_{\text{mag}}^i$ \cite{Menezes:2008qt.Menezes:2009uc}.

\section{Results}

In the present section we investigate the effect of the magnetic field on the 
strange quark chiral phase transition. The strange quark is strongly coupled to 
the light quarks through the 't Hooft term. 
It has already been shown that a $U_A(1)$ anomaly reduction in the medium 
has the effect of weakening the chiral phase transition \cite{Fukushima:2008wg}.
Therefore, understanding the strange quark chiral phase transition 
requires that the effect of the 't Hooft term is understood. 
Another aspect already mentioned is its high mass when compared with the light quarks. 
In order to feel an effect similar to the $d$ quark in the presence of a magnetic 
field, a much stronger field must be applied. 

In the following we analyze the effect of the 't Hooft term, the
current $s$ quark mass and the inverse magnetic catalysis effect that
can be taken into account by setting a magnetic field dependent 
scalar coupling \cite{Ferreira:2014kpa}.

\begin{figure}[t]
    \includegraphics[width=0.9\linewidth,angle=0]{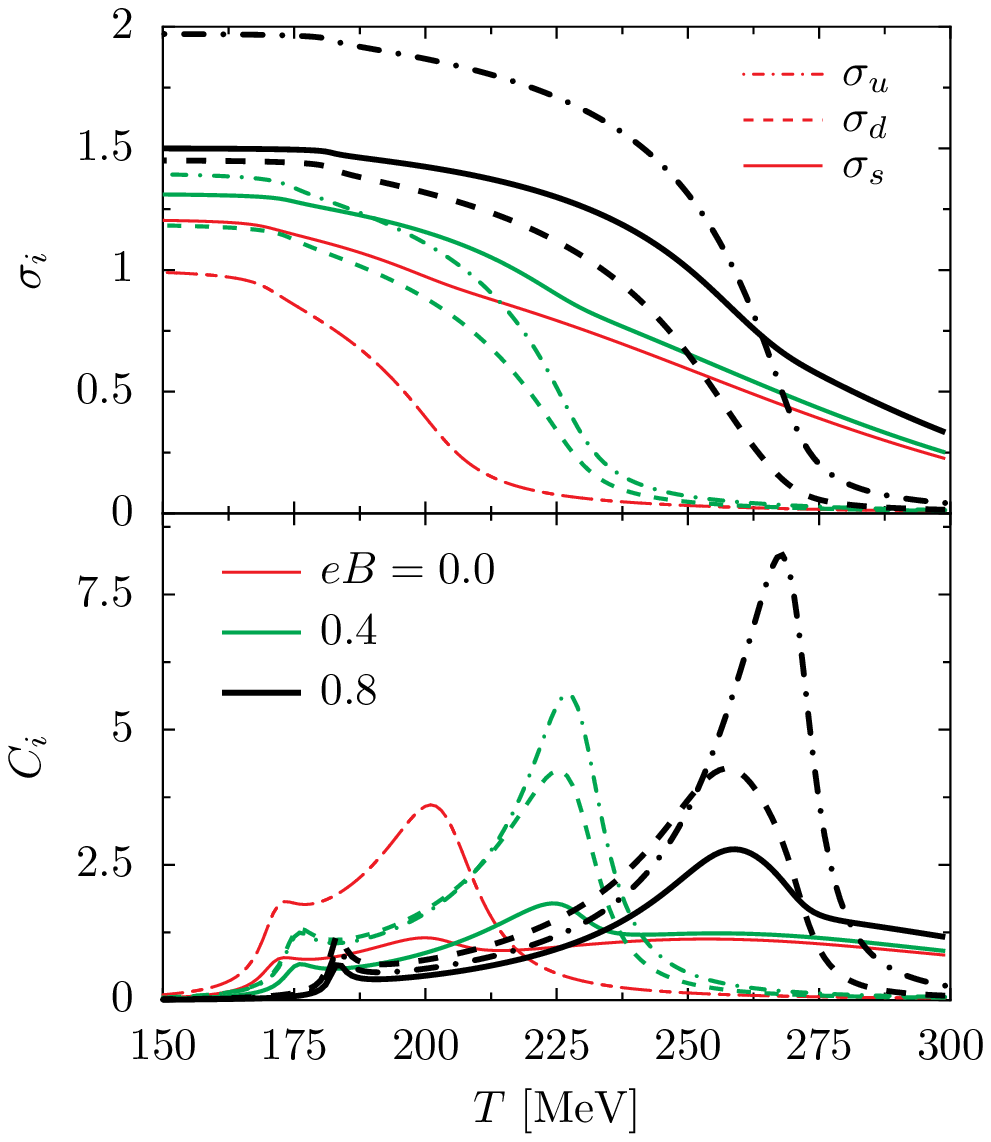}
    \caption{The quark condensates and their susceptibilities 
    as a function of temperature for three magnetic field strengths: 
    $eB=0.0,0.4$ and $0.8$ GeV$^2$.}
\label{fig:condensados}
\end{figure}

We first analyze the order parameters within the complete PNJL model,
including the  't Hooft term. 
In Fig. \ref{fig:condensados} the normalized quark condensates 
$\sigma_i=\ev{\bar{q}_iq_i}(B,T)/\ev{\bar{q}_uq_u}(0,0)$   and their
respective susceptibilities $C_i = -m_\pi\partial\sigma_i/\partial T$
are plotted for three magnetic field
strengths. 
The quark condensates are normalized by the up quark vacuum condensate 
value at zero magnetic field, and the inclusion of the pion mass
$m_\pi$ in the susceptibilities  assures a dimensionless quantity.
The quark condensates are enhanced by the presence of the magnetic
field, effect known by magnetic catalysis. 
For $eB=0.8$ GeV$^2$, due to the quark electric charge difference, 
the condensate $\sigma_u$ is  larger than $\sigma_s$,  even though the 
current strange quark current mass $m_s$ is larger than the current 
$u$ quark mass $m_u$. The first peaks in the susceptibilities at low 
temperatures are induced by the deconfinement transition, i.e. by the 
rapid change of the Polyakov loop with temperature, that signals the 
deconfinement phase transition. 
As already pointed out in \cite{Ferreira:2013tba}, compared with the 
chiral transition, the deconfinement phase transition is quite 
insensitive to the presence of the magnetic field.

\subsection{The impact of the 't Hooft term on the chiral phase transitions}

\begin{figure}[t]
    \includegraphics[width=0.9\linewidth,angle=0]{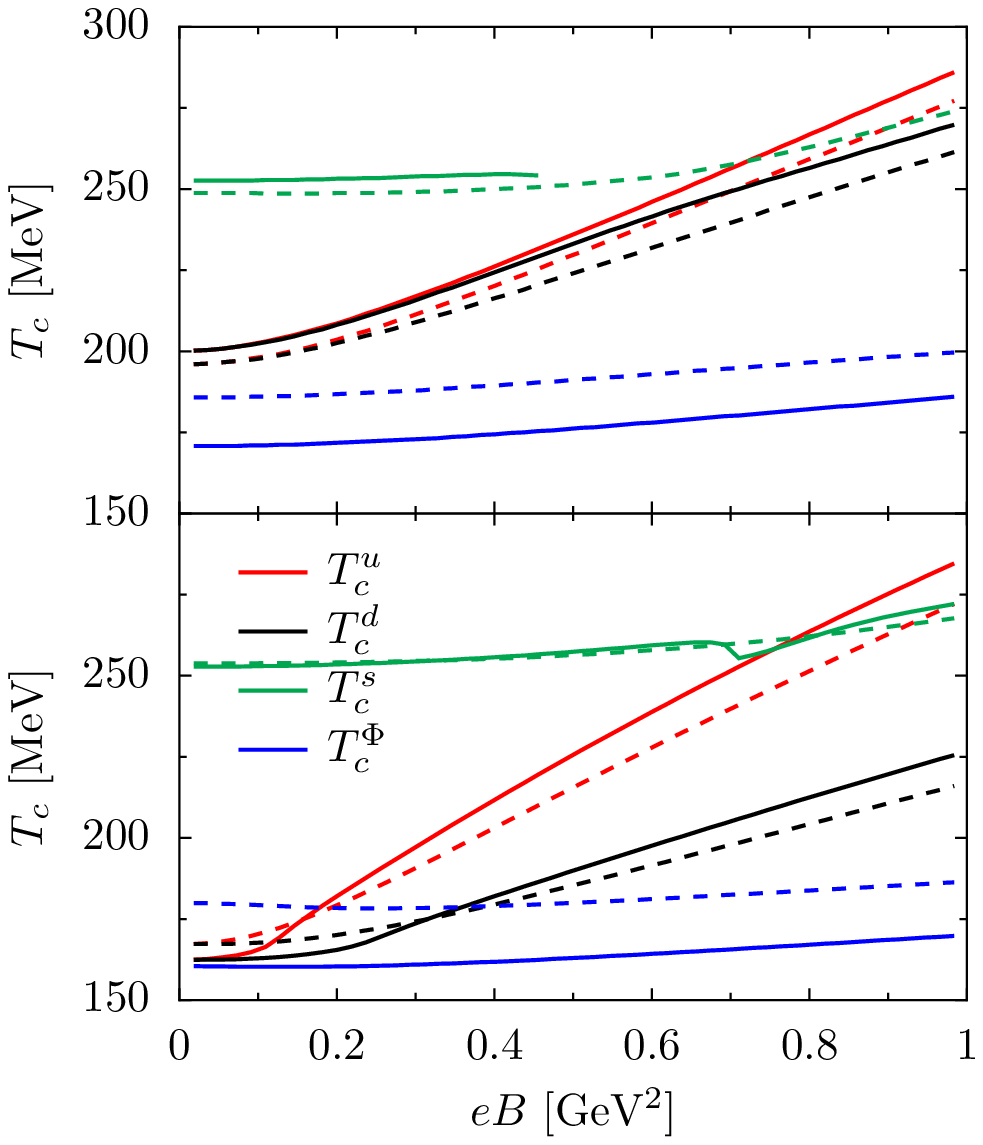}
    \caption{The critical temperatures ($T_c^i$) as a function of $eB$, 
    for $K\neq0$ (top panel) and $K=0$ (bottom panel),
    using the peak of the susceptibilities (solid lines)
    and half the vacuum value of the order parameters (dashed lines).}
\label{fig:Tcs_Gd_0}
\end{figure}

In order to analyze the impact of the 't Hooft term on the transition temperatures 
of the chiral transition as a function of the magnetic field strength, we calculate 
the transition temperatures with $K\neq0$ and $K=0$.
The results are shown in Fig. \ref{fig:Tcs_Gd_0}. 
Two criteria are used to calculate the transition temperatures: 
(i) the peaks of the respective susceptibilities and (ii) the temperature $T_c^i$
at which the order parameter is half the respective vacuum value,  
$\ev{\bar{q}_iq_i}(B,T_c^i)=0.5\ev{\bar{q}_iq_i}(B,0)$. 
For the $K\neq0$ case (top panel), using the first criteria (solid lines), the critical 
temperature for the strange quark can only be calculated up to some maximum 
$eB$ value. For higher $eB$ values, the chiral transition for the $u$ and $d$ quarks 
washes out the strange quark transition and the inflection point of the strange quark 
condensate, which defines  the strange quark phase transition, cannot be defined anymore. 
This can be solved if the second criterion (dashed lines) is used. 
With the second criteria,  a similar behavior is obtained for the $s$ quark, 
but with lower transition temperatures. 
From the top panel of Fig. \ref{fig:Tcs_Gd_0} it is also seen that the transition 
temperatures for the light quarks increase faster with $eB$  than for the $s$ quark. 
In fact, the strange transition temperature is almost insensitive to the 
magnetic field strength up to $eB\approx0.4$ GeV$^2$, mainly due to its larger mass.
Another interesting aspect that can be seen is the increase of the splitting
between the temperatures at which chiral transitions occur, for the light quarks, and 
the deconfinement temperature. This particular feature was already found in the context 
of the linear sigma model coupled to quarks and to the Polyakov loop in \cite{Mizher:2010zb}. 
The Sakai-Sugimoto model also predicts a similar behavior \cite{Callebaut:2013ria}.

Comparing the $K=0$,  Fig. \ref{fig:Tcs_Gd_0} bottom panel, with the
$K\neq0$ case,  Fig. \ref{fig:Tcs_Gd_0} top panel, some important
features should be pointed out in the light quark sector:
(1) for low $eB$ values, smaller chiral transition temperatures are obtained, 
and the difference $T_c^u-T_c^d$ increases faster with $eB$; 
(2) for low $eB$ values, and using the second criterion for the transition temperature,
the deconfinement transition temperature lies above the chiral 
transitions temperatures as obtained in LQCD calculations;
(3) at $eB\approx0$, the gap between the chiral and deconfinement transitions is quite small. 
We conclude that the quark chiral transitions and the deconfinement transition are
strongly correlated due to the 't Hooft term, and some features of the
QCD phase diagram are precisely defined by this term.

\begin{figure}[t]
    \includegraphics[width=0.9\linewidth,angle=0]{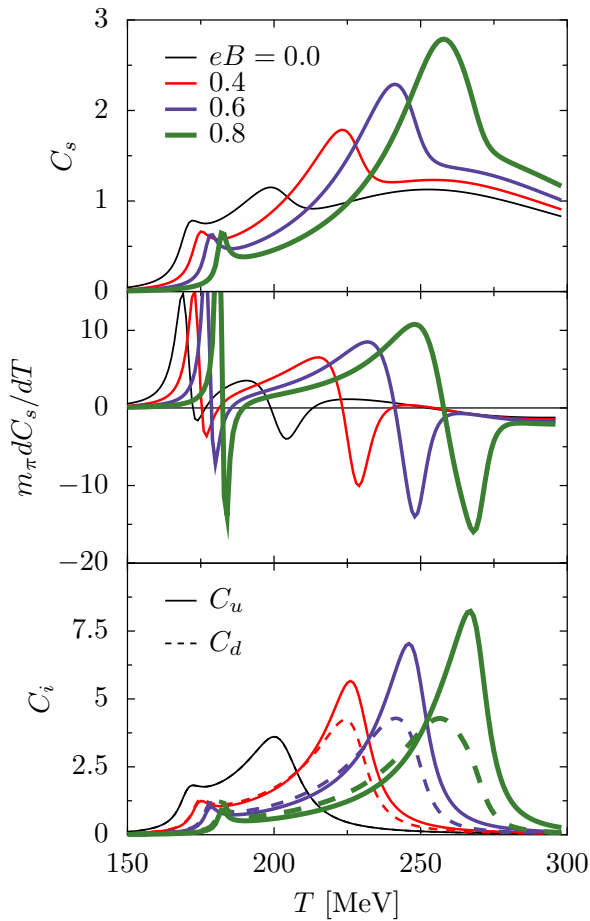}
    \caption{(Top panel) The strange quark susceptibilities $C_s$, (middle panel)
    $dC_s/dT$ and (bottom panel) the $u$ (solid lines) and $d$ (dashed lines) quark
    susceptibilities $C_{u,d}$ as a function of $eB$  with the 't Hooft term. }
\label{fig:cond_s_with}
\end{figure}
\begin{figure}[t]
    \includegraphics[width=0.9\linewidth,angle=0]{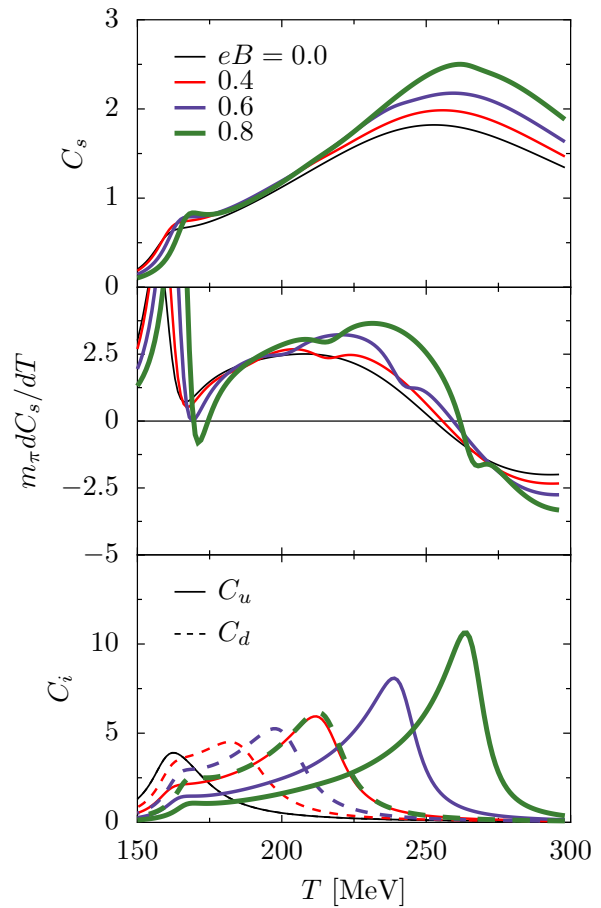}
    \caption{(Top panel) The strange quark susceptibilities $C_s$, (middle panel)
    $dC_s/dT$ and (bottom panel) the $u$ (solid lines) and $d$ (dashed lines) quark 
    susceptibilities $C_{u,d}$ as a function of $eB$ without the 't Hooft term. }
\label{fig:cond_s_without}
\end{figure}

In order to understand how the magnetic fields affect the strange quark 
and, thus, its critical temperature, it is important to determine the impact that 
the chiral restoration of the light sector has on the behavior
of the strange quark condensate. In Figs. \ref{fig:cond_s_with} and \ref{fig:cond_s_without} we
calculate for $K\neq0$ and $K=0$, respectively, the strange quark susceptibilities 
$C_s$ (top panel), the derivative of the susceptibilities $m_{\pi}dC_s/dT$ (middle panel), 
and the susceptibilities of the light quarks $C_{u,d}$ (bottom panel), for several values of $eB$.
For $K\neq0$, the strange quark transition is influenced more strongly by the chiral restoration 
of the light sector than for $K=0$. 
In fact, with $K\neq0$, the most pronounced peek in the strange quark susceptibility $C_s$ is due 
to the chiral transition of the $u$ and $d$ quarks (see Fig. \ref{fig:cond_s_with} bottom panel),
because with a finite 't Hooft term the gap equations mix all flavors.

The strange quark transition is reflected in the last inflection point of $C_s$.
In the middle panel of  Fig. \ref{fig:cond_s_with}, it is seen that for $eB=0.6$ GeV$^2$ 
this inflection point disappears, being washed out by the transition of the light quarks. 
With $K=0$ there is no flavor mixing in the gap equations, and the 
strange quark phase transition is clearly identified in the susceptibility. 
Although some bumps still appear in the derivative of the $C_s$ due
to the light quarks, their intensity is much weaker than the transition
of the strange quark itself. The mixing occurs through the
distribution functions.
The no flavor mixing for $K=0$ is confirmed  in the bottom panel of Fig. 
\ref{fig:cond_s_without}, where  it is seen that there is no direct coupling 
between the $u$ and $d$ quarks as the magnetic field increases.

\subsection{The impact of the current strange quark mass on its transition temperature}

As we have seen in the last section, the restoration of the chiral symmetry of 
the strange quark has a different behavior when compared with the light quarks
due to its larger current mass: $m_s\approx25.5m_{u,d}$. 
For low magnetic field strengths, the critical temperature of the strange quark  
does not change much as compared with the light quarks. 
As expected, the restoration of the chiral symmetry will depend not only on the 
quark electric charges but also on their current quark masses. 
Effects of the magnetic field become noticeable when $eB$ becomes of the order 
of the quark mass squared.

Next we will analyze how the restoration of the chiral symmetry depends
on the value of the strange quark current mass $m_s$, keeping $m_{u,d}=5.5$ MeV.
In this section the PNJL model with 't Hooft term ($K=12.36/\Lambda^5$) will
be used. In NJL and PNJL models, at $eB=0$, this dependence was investigated in
\cite{costa}. 
 
\begin{figure}[t]
    \includegraphics[width=0.9\linewidth,angle=0]{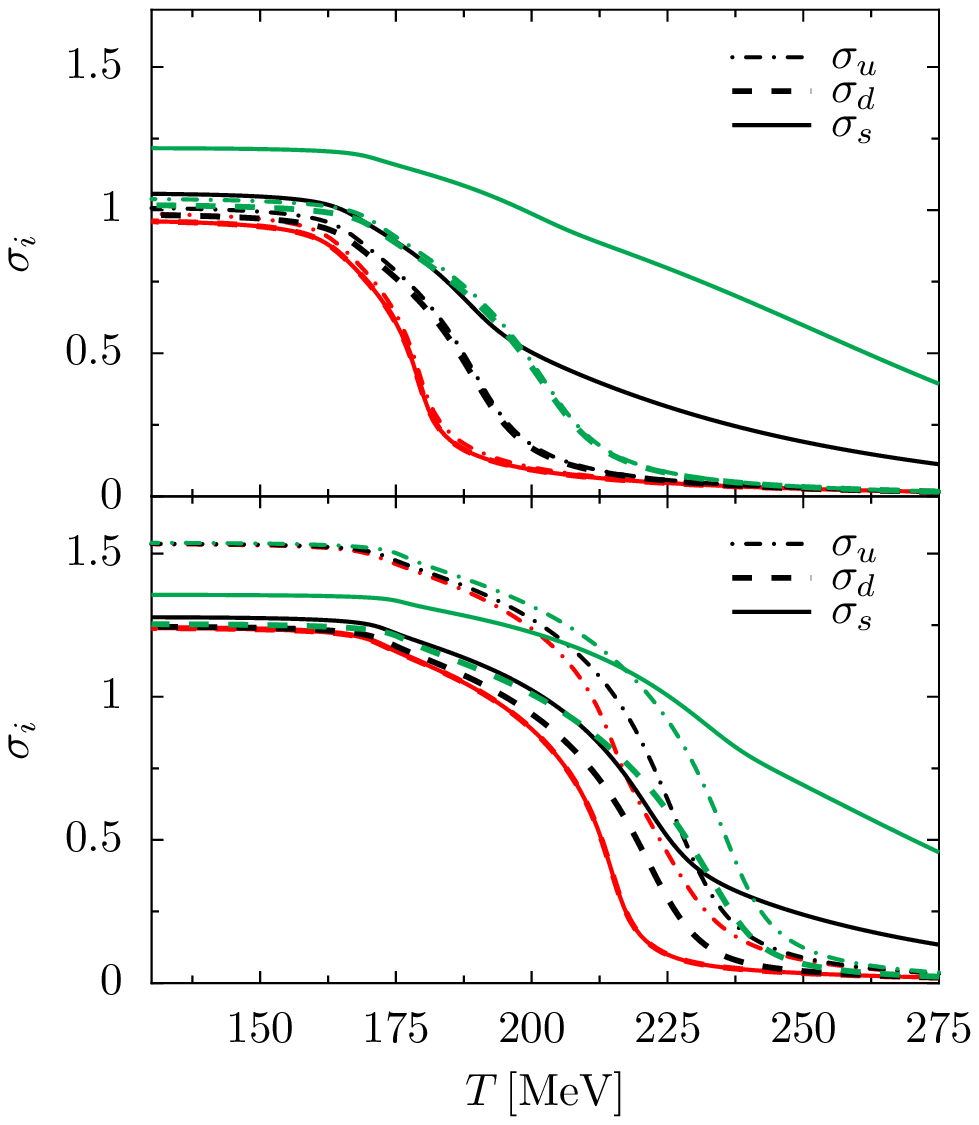}
    \caption{The order parameters as a function of temperature for $m_s=m_{u,d}=5.5$ MeV
     (red lines), $m_s=40$ MeV (black lines), and $m_s=140.7$ MeV (green lines) for $eB=0.1$ GeV$^2$ 
     (top panel) and $0.5$ GeV$^2$ (bottom panel).}
\label{fig:fig_condensados_ms}
\end{figure}

We first investigate the impact of the current mass of the strange quark on the quark condensates.
In Fig. \ref{fig:fig_condensados_ms} the renormalized quark condensates are plotted
as function of temperature, for $eB=0.1$ GeV$^2$ (top panel) and $eB=0.5$ 
GeV$^2$ (bottom panel), using three values of strange current mass:
$m_s=m_{u,d}=5.5$ MeV, $40$ MeV, and $140.7$ MeV. We have
renormalized the condensates as $\sigma_i(B,T)=\sigma_i(B,T)/\sigma_u(0,0)$,
where $\sigma_u(0,0)$ is the vacuum condensate at zero magnetic field, with 
the current mass of the $u$ quark.
For $m_s=m_{u,d}$, the three quarks form an isospin triplet that is broken by
the magnetic field presence. Therefore, the differences in the 
condensates are only induced by the electric charge of each quark,
having the $\sigma_u$ as the highest value ($|q_u|=2e/3$), and both $\sigma_{d}$ and $\sigma_{s}$ 
the lowest ($|q_{d,s}|=e/3$).  The effect of the charge is always
present independently of the quark masses. 

The degeneracy of both $\sigma_{d}$ and $\sigma_{s}$ is lifted when we set $m_{u,d}\neq m_s$, 
that is, for $m_s=40$ and $140.7$ MeV. We see that for a low magnetic field strength, 
$0.1$ GeV$^2$ (Fig. \ref{fig:fig_condensados_ms} top panel), for $m_s=m_{u,d}$ (red), the 
$u$ condensate due to its electric charge has the highest value at any temperature. 
However, if  $m_s=40$ (black line) and $140.7$ MeV (green line), 
the $s$ quark condensate has the highest value. 
At low $eB$ the magnetic catalysis effect is mainly determined by the charge of the quark if 
$m_s$ is of the order of $m_{u,d}$, but this effect becomes weaker with increasing mass $m_s$. 
As the strange current quark mass increases the restoration of chiral symmetry in the light 
sector is pushed to higher temperatures, due to the flavor mixing induced by the 't Hooft term.

For larger magnetic fields, i.e.  $0.5$ GeV$^2$ (Fig. \ref{fig:fig_condensados_ms} bottom panel) 
and at low temperatures, the $u$ and $d$ quark condensates are not much affected by the $m_s$ value. 
The effect of the quark electric charge in the magnetic catalysis at low temperatures 
predominates over the effect of the strange current quark mass.

\begin{figure}[t]
    \includegraphics[width=0.9\linewidth,angle=0]{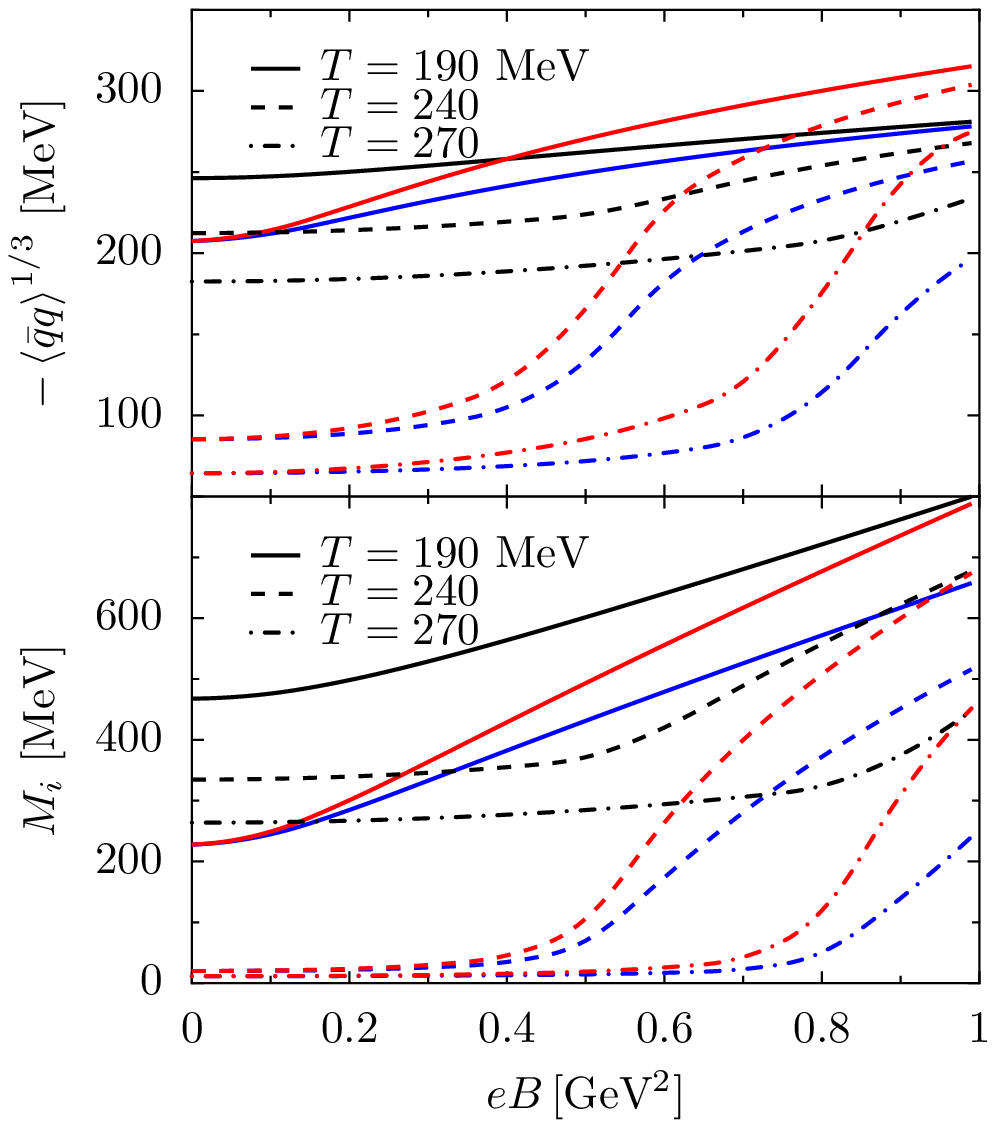}
		\caption{The condensates (top panel) and masses (bottom panel) of the $s$ (black lines), 
		$u$ (red lines), and $d$ (blues) quarks as a function of $eB$
		for several temperatures ($m_s=140.7$ MeV).}
\label{fig:cond_massas}
\end{figure}

\begin{figure*}[t]
    \includegraphics[width=0.4\linewidth,angle=0]{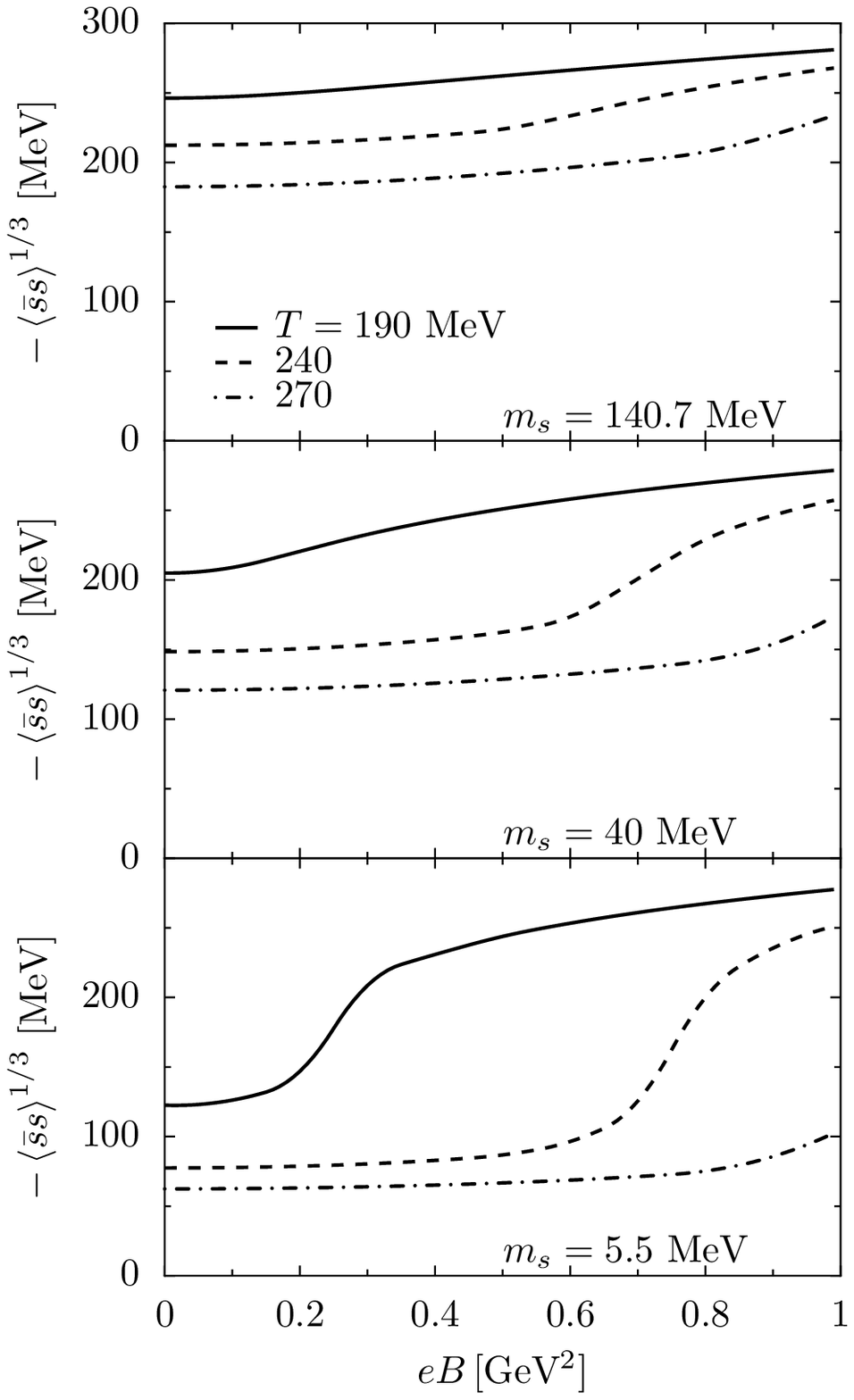}\hspace{1.cm}
    \includegraphics[width=0.4\linewidth,angle=0]{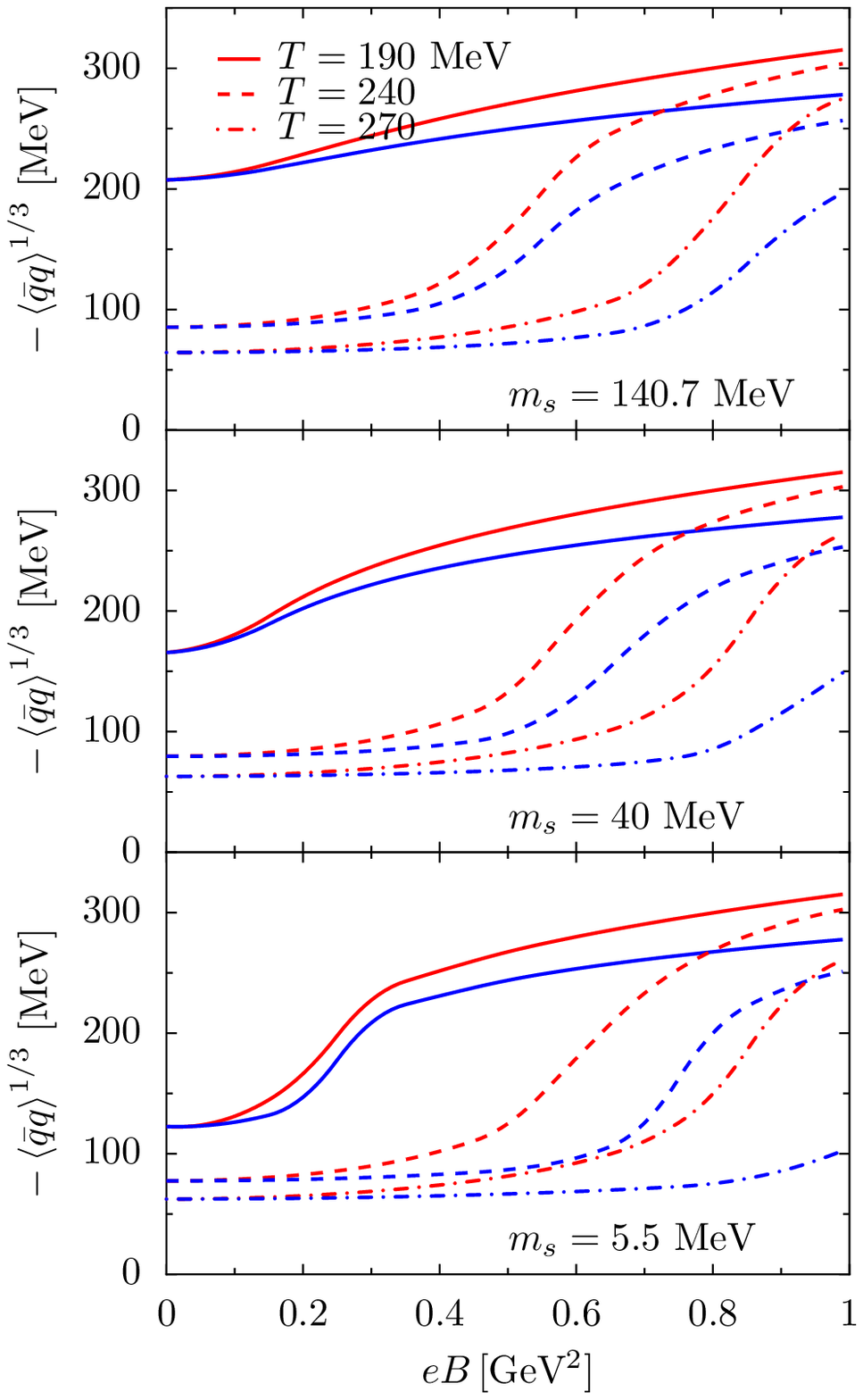}
    \caption{The quark condensates: $s$ (left), $u$ (red lines) and $d$ (blue lines) (right panel) 
    for several temperatures  for three current strange quark mass values: 
    $140.7$ MeV (top panels), $40$ MeV (middle panels), and $5.5$ MeV (bottom panels).}
\label{fig:cond_ms}
\end{figure*}

In Fig. \ref{fig:cond_massas}, we fix the $m_s$ value to its current mass of $140.7$ MeV, 
and calculate the quark condensates (top panel) and masses (bottom panel) as a function of $eB$, 
for three temperature values. The different behavior between both 
sectors is clear: for $T=240$ and $270$ MeV, and at low $eB$, the light quarks 
are in a restored chiral phase, but at some higher value of $eB$, the magnetic field 
drives the light quarks into a chiral broken phase, manifested in the sudden 
increase of the condensate values.
This occurs at larger values of $eB$ for larger temperatures.
The values of the strange quark condensate and mass are high for all
the magnetic field intensity range shown, and for the three temperatures. 
Although it is difficult to define the chiral 
restored/broken phase for the strange quark, we can see a similar behavior as in the 
light sector, mainly from the bottom panel with the quark masses at $T=240$ and 
$270$ MeV: the strange quark condensate increases slightly with $eB$ for low magnetic 
fields and  at some value of $eB$ there is a steeper increase of the masses.

In Fig. \ref{fig:cond_ms} we perform the same calculation as we did in Fig. \ref{fig:cond_massas}, 
but now for three $m_s$ values: $5.5$ MeV, $40$ MeV, and $140.7$ MeV. 
In the bottom panel we have three degenerate quark masses and, as said before, 
the differences between the different flavors are only due to the quark electric charge. 
As $m_s$ increases, in the center and top of Fig. \ref{fig:cond_ms} (left), the strange quark 
condensate gets less affected by $eB$, reflecting its higher constituent mass and the 
consequent shift  of the chiral restoration to larger temperatures. As can be seen 
on the right panel of Fig. \ref{fig:cond_ms}, the light sector  also feels the change in 
$m_s$. This is more clearly seen for  $T=190$ MeV: in this case  the condensates soften with 
increasing $m_s$. As $m_s$ increases its value, due to the flavor mixing, not only the critical 
transition temperature of the strange quark increases, but also the transition of the light 
quarks is shifted to larger temperatures.

\begin{figure}[t]
    \includegraphics[width=0.9\linewidth,angle=0]{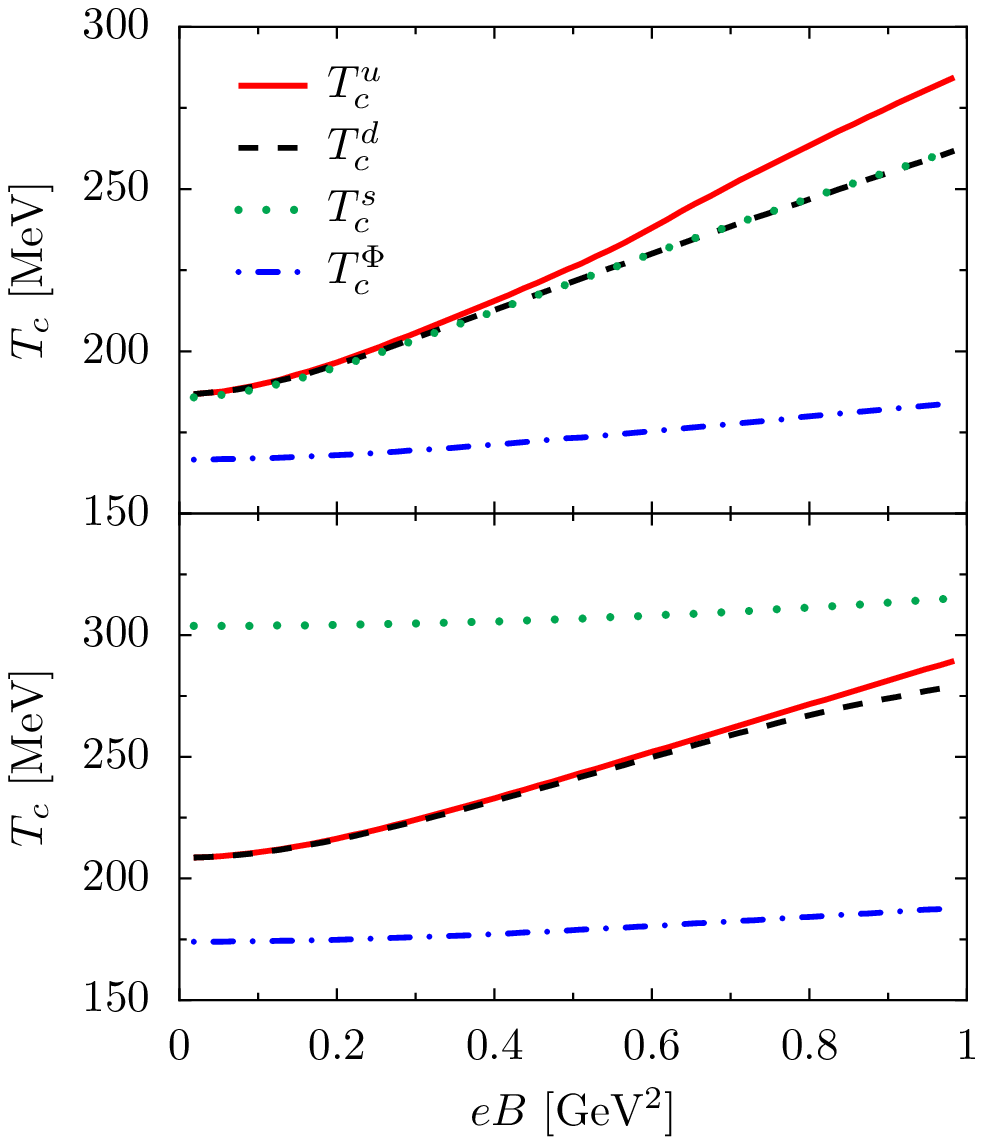}
    \caption{The critical temperatures for two values of the current quark mass as a function of $eB$: 
    $m_s=40$ MeV (top panel) and $m_s=300$ MeV (bottom panel).}
\label{fig:Tcs_ms}
\end{figure}

We will next calculate the critical temperatures as a function of $eB$ for two cases: 
an intermediate case between the light and heavy quark sectors, $m_s=40$ MeV,
and an extreme heavy case, $m_s=300$ MeV. The result is presented in Fig. \ref{fig:Tcs_ms}. 
Two main conclusions can be drawn: 
(1) for $m_s=40$ MeV and at high magnetic fields ($eB>0.3$ GeV$^2$), the transition 
of the strange quark occurs at the same temperature as the $d$ quark. 
This indicates that at sufficiently high magnetic field, the critical temperatures 
at which the chiral symmetry restoration occurs are mainly determined by the electric 
charge of the quark, because the current quark masses of all  quarks are not too different; 
(2) for $m_s=300$ MeV, the critical temperature of the strange quark does not change much 
with the magnetic field due to its very large mass. 

Although the magnetic catalysis affects all quarks, the light sector shows an increase 
of the critical temperature with the magnetic field while  the strange sector is almost 
insensitive at low magnetic fields, and increases slightly for high magnetic fields.

\subsection{Inverse magnetic catalysis}

\begin{figure}[t]
\includegraphics[width=0.9\linewidth,angle=0]{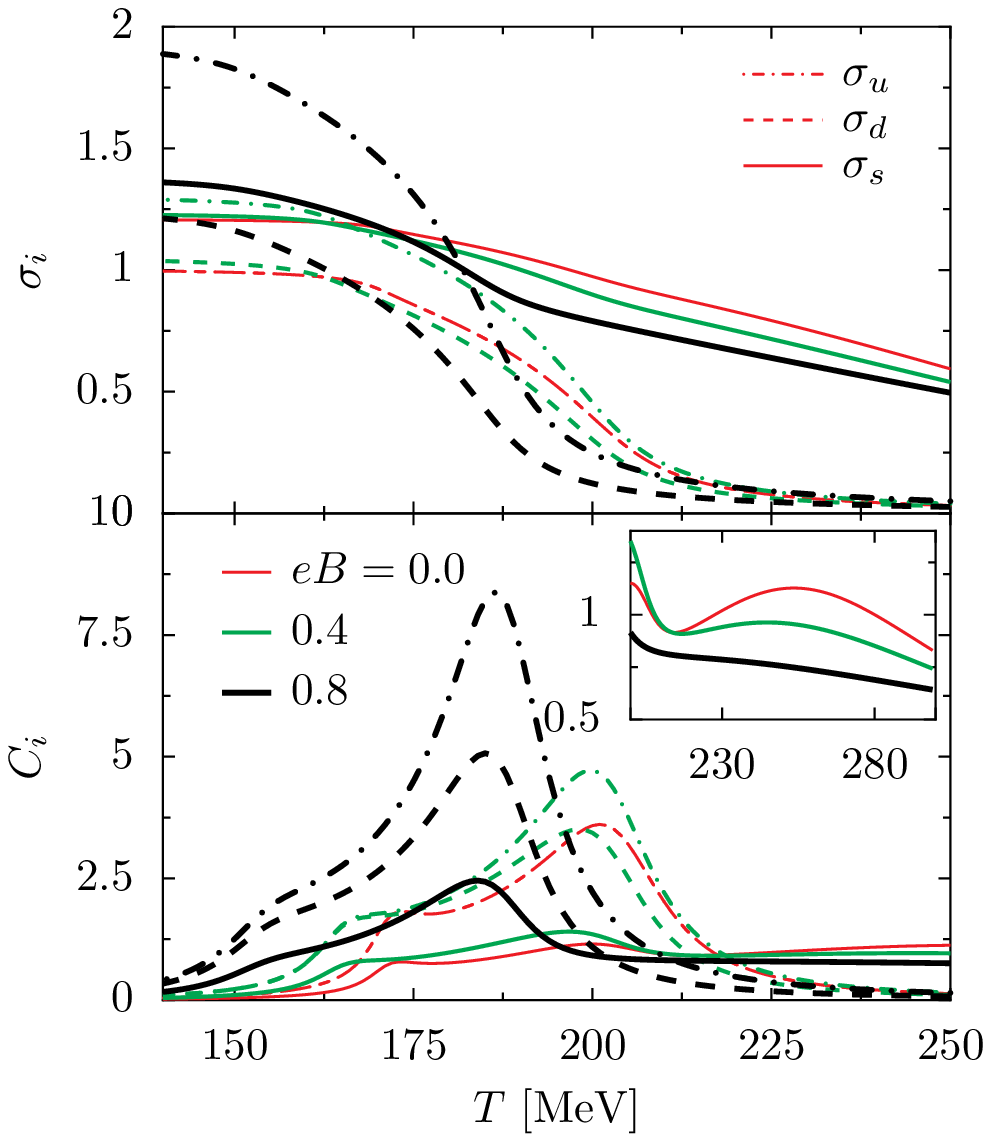}
\caption{The quark condensates and their susceptibilities 
    as a function of temperature for
    $eB=0.0,0.4$ and $0.8$ GeV$^2$, using $G_s(eB)$.}
\label{fig:fig9}
\end{figure}

\begin{figure}[t]
\includegraphics[width=0.9\linewidth,angle=0]{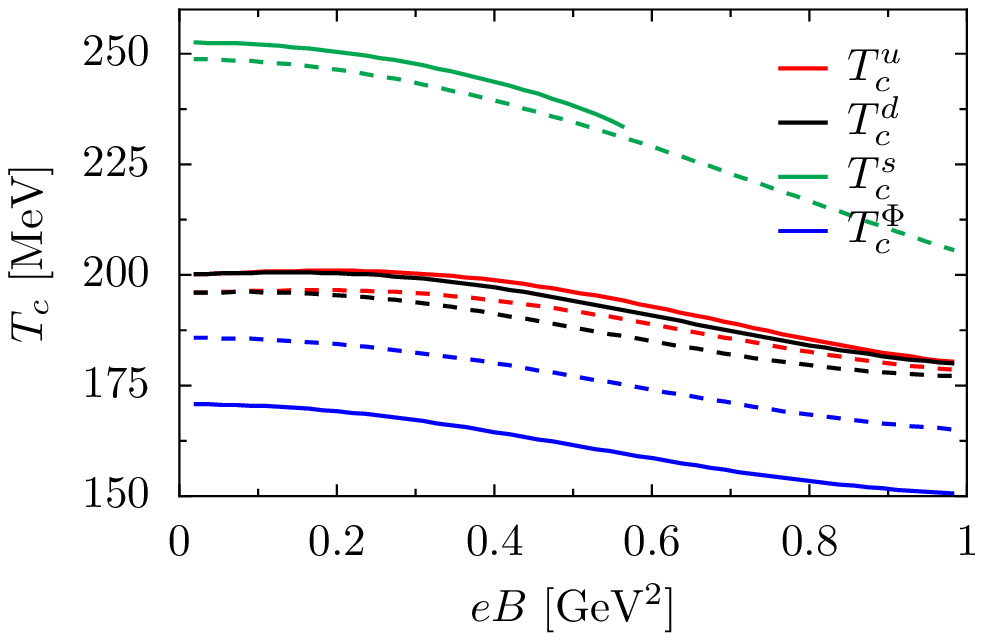}
\caption{
		The critical temperatures $T^i_c$ as a function of $eB$,
    given by the peak of the susceptibilities (solid lines)
    and half the vacuum value of the order parameters (dashed lines),
    using $G_s(eB)$.}
\label{fig:fig10}
\end{figure}

LQCD results show that the chiral and the deconfinement transition 
decreases with the magnetic field \cite{Bali:2011qj}. This is due to the inverse magnetic 
catalysis mechanism, which suppresses the condensates near the crossover 
transition and favors higher values for the Polyakov loop. Within effective 
models, and, in particular, in the PNJL model, there have been several attempts to 
reproduce the inverse magnetic catalysis mechanism 
\cite{Ferreira:2013tba,Farias:2014eca,Ferreira:2014kpa}. 
In \cite{Ferreira:2014kpa}, a magnetic field dependent coupling $G_s(eB)$ was introduced: 
this dependence was fitted to reproduce, within the NJL model, the renormalized critical 
temperature of the chiral transition given by LQCD \cite{Bali:2011qj}.
Therefore, it is interesting to study the strange quark transition 
using this approach. We take  $G_s(eB)$ as defined in \cite{Ferreira:2014kpa}
\begin{equation}
G_s(\zeta)=G_s^0\pc{\frac{1+a\,\zeta^2+b\,\zeta^3}
{1+c\,\zeta^2+d\,\zeta^4}}\,
\label{eq:fit}
\end{equation}
where $a = 0.0108805$, $b=-1.0133\times10^{-4}$, $c= 0.02228$,  $d=1.84558\times10^{-4}$ and
$\zeta=eB/\Lambda_{QCD}^2$ with $\Lambda_{QCD}=300$ MeV.
 
We have plotted in Fig. \ref{fig:fig9} the quark condensates and 
the susceptibilities using  $G_s(eB)$.
In Fig. \ref{fig:fig10} the critical temperatures are plotted as 
a function of $eB$. All critical temperatures decrease with $eB$. 
Looking at the condensates behavior, we see that all of them are 
enhanced at low temperatures, suppressed at temperatures near the 
transition temperature and enhanced again at high temperatures. 
Also the first peaks in the susceptibilities, induced by the deconfinement 
transition, are shifted to lower temperatures with increasing $eB$. 
A larger $m_s$ would still give rise to the same kind of effects but
less pronounced.

In order  to understand why the critical temperature of the strange quark can only be 
defined up to a certain $eB$ (using the first criteria), we show in the lower panel of 
Fig. \ref{fig:fig9} a zoom of the strange transition region. 
The maximum of the strange susceptibility  induced by the chiral transition of the 
strange quark is washed out for larger $eB$ values.  

\begin{figure}[t]
\includegraphics[width=0.8\linewidth,angle=0]{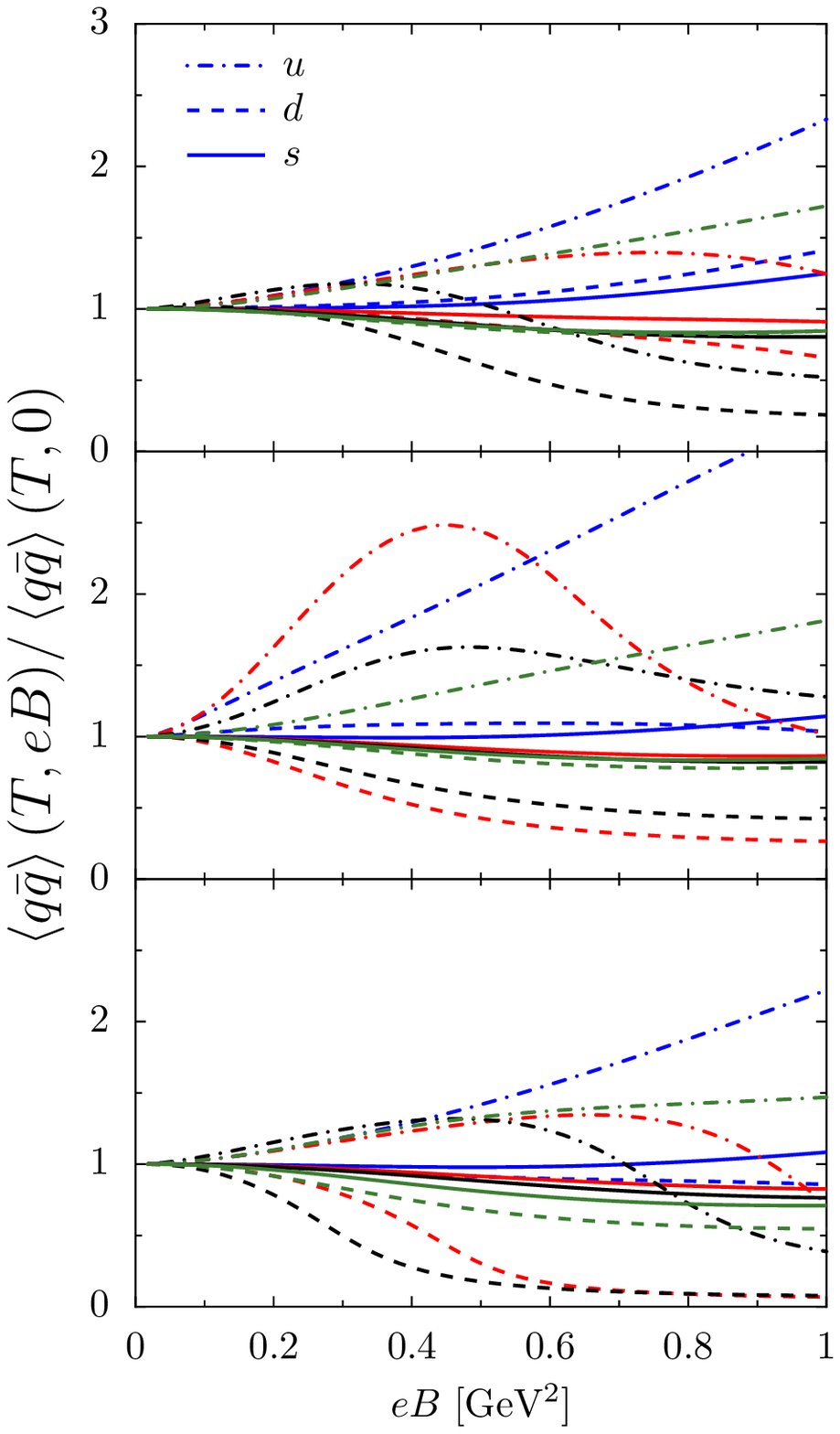}
\caption{
				The ratios of the $u$, $d$ and $s$ condensates, 
				$\langle  q_i \bar q_i \rangle (T, eB)/\langle q_i \bar q_i \rangle (T, 0)$, 
				as a function of $eB$, for several values of $T$ [0 (blue lines), 180
        (red lines), 200 (black lines) and 250 (green lines) MeV] using $G_s(eB)$: 
        upper panel including the 't Hooft term; 
        middle panel excluding the 't Hooft term without refitting the other parameters;
        and bottom panel excluding the 't Hooft term and using the
        parametrization of Ref. \cite{Costa:2005}.
        }
\label{fig:fig11a}
\end{figure}

The IMC effect is strongly influenced by the 't Hooft term as we
will show in the following. In Fig. \ref{fig:fig11a} the $u$, $d$ 
and $s$ condensates normalized to their values for a zero magnetic 
field are plotted  for a magnetic field dependent coupling $G_s(eB)$ 
and different scenarios for the 't Hooft term: 
with the 't Hooft term (upper panel); 
with the 't Hooft term switched off, but no refitting of the couplings in 
order to reproduce the vacuum properties of the pion and kaon (middle panel); 
we switch off the 't Hooft term and use the parametrization proposed in 
\cite{Costa:2005} that reproduces the pion and kaon properties without the 't Hooft term
(bottom panel).

Just like the $u$ and $d$ quarks, the $s$ quark also shows the inverse magnetic 
catalysis effect: see in Fig. \ref{fig:fig11a} (upper panel) the dashed-dotted lines. 
The strange quark condensate presents a nonmonotonic behavior as a 
function of $eB$, and its critical temperature is a decreasing function of $eB$. 
This behavior is not following the trend indicated in \cite{Bali:2012zg}, where the 
$s$ quark condensate is said to increase with growing $B$ for all temperatures. 
The IMC effect is still present if the $K$ is set to zero: see Fig. \ref{fig:fig11a}
(middle panel). In this case the $d$ and $s$ quarks fill a stronger IMC effect, 
because the mixing with the $u$ quark, with a much larger magnetic catalysis effect 
due to its larger charge, which prevents a fast decrease of the other condensates, does not exist. 
The results of Fig. \ref{fig:fig11a} (bottom panel) were also obtained excluding the 
't Hooft term, but using a different parametrization that describes the vacuum
properties of the pion and kaon \cite{Costa:2005}. The general behavior is similar 
to the results shown in the middle panel, although the $u$ quark shows a behavior closer 
to the upper panel, were the 't Hooft term was included.
This is due to the larger mass of the $u$ quark within parametrization \cite{Costa:2005}, 
that compensates the effect of the strong magnetic field due to its higher charge.

It is worth pointing out that the behavior of the $s$ quark condensate
is expectable. Being the quark with larger mass, it will not feel
a strong magnetic catalysis  for weak fields. Moreover, its charge is
half of the $u$ quark charge, and, therefore, it is also not as affected
as the $u$ quark. On the other hand, the IMC effect is implemented in the 
present model through a parametrization of the scalar coupling, and, 
consequently, is switched on as soon as $eB>0$. 
From these two effects, it  results that the $s$ quark also feels the IMC effect.
This is clearly seen by switching off the 't Hooft term. In this case no mixing with the $u$ 
quark occurs and the $s$ condensate decreases for low $eB$ even for $T=0$.

\subsection{Thermodynamical properties}

\begin{figure*}[t]
\begin{tabular}{ccc}
\includegraphics[width=0.32\linewidth,angle=0]{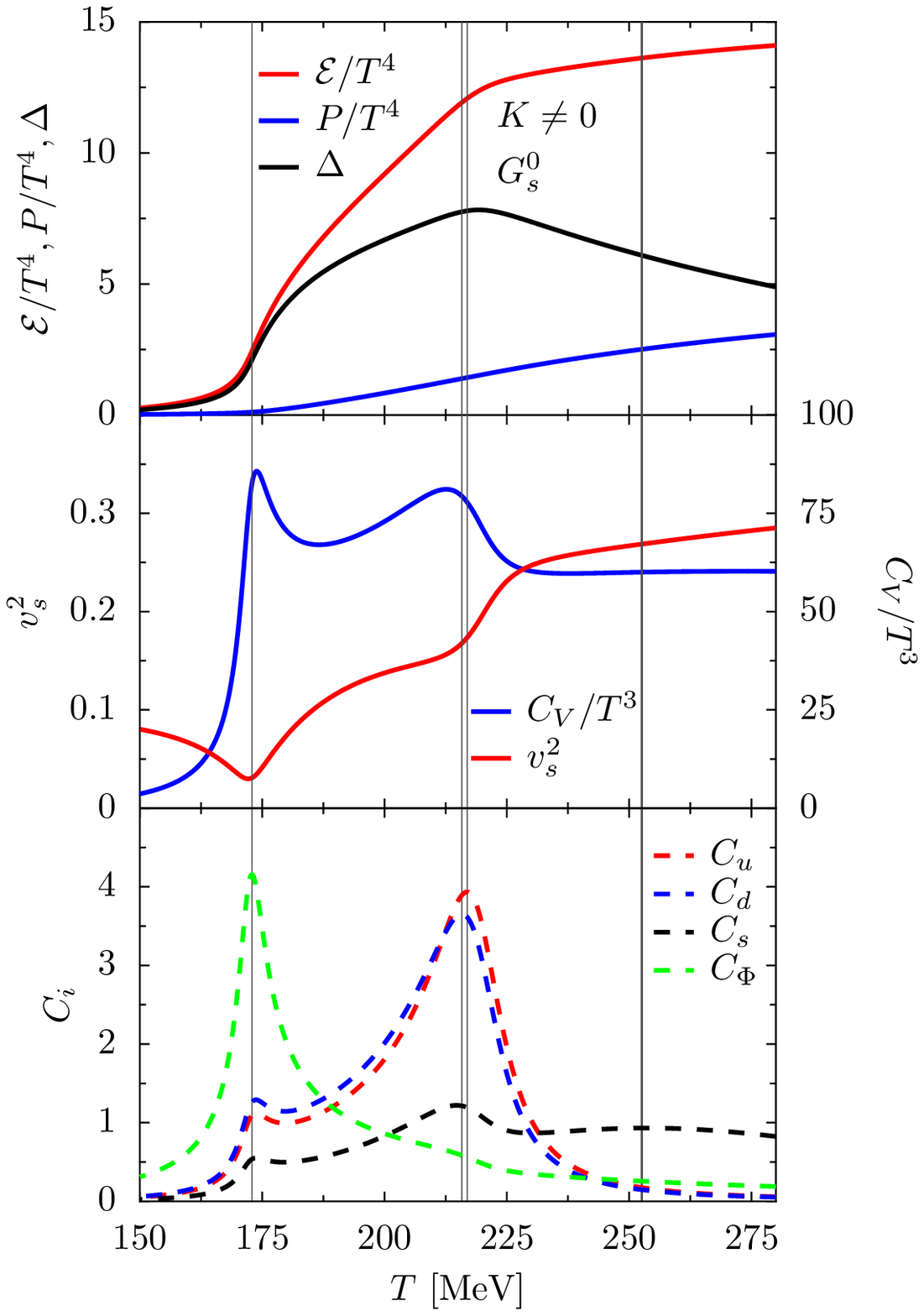}&
\includegraphics[width=0.32\linewidth,angle=0]{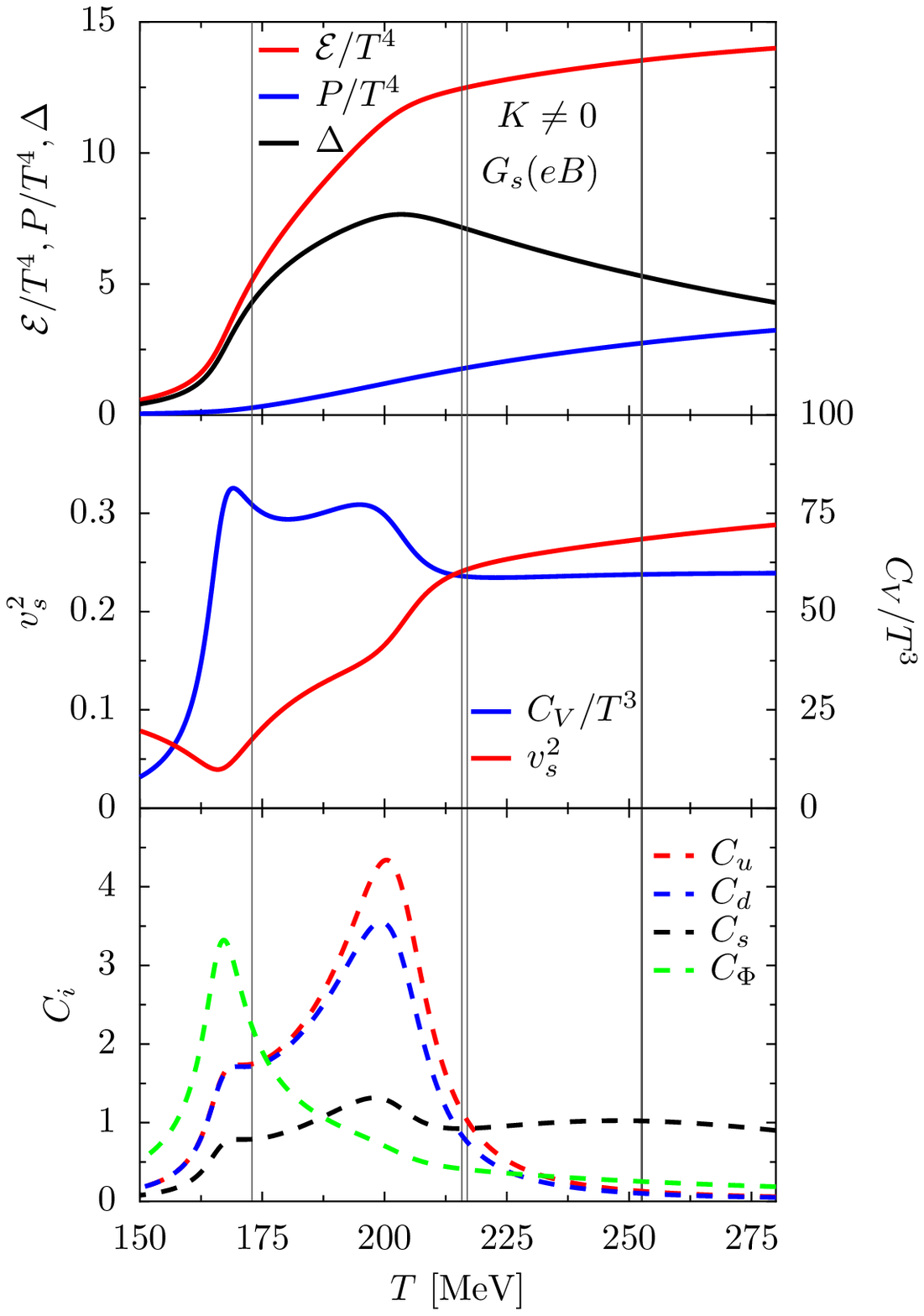}&
\includegraphics[width=0.32\linewidth,angle=0]{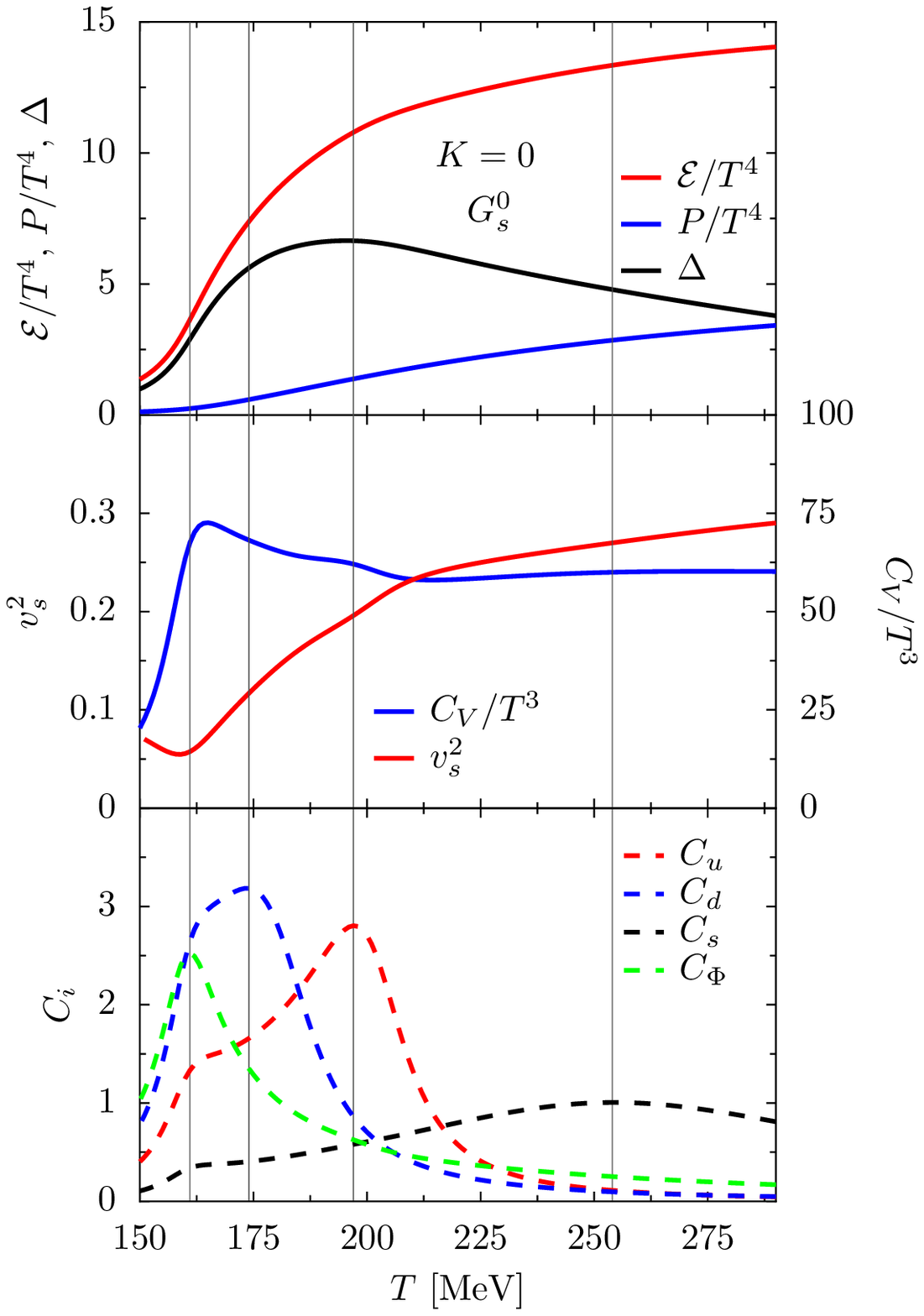}\\
\end{tabular}
    \caption{The scaled energy density ${\cal E}/T^4$, the interaction measure 
		$\Delta(T)=({\cal E}-3P)/T^4$, and the scaled pressure $P/T^4$ as a function
		of temperature $T$ (top panels); the scaled specific heat $C_V/T^3$, and speed of 
		sound squared $v_s^2$ as a function of temperature $T$ (middle panels); and the 
		quark susceptibilities (bottom panels), for $eB=0.3$ GeV$^2$ with the 't Hooft term, and
		$G_s=G_s^0$ (left) or  $G_s=G_s(eB)$ (middle) and without the 't Hooft term and $G_s=G_s^0$ (right). 
		The vertical lines in the left and right panels indicate the position of the maximum of the 
		quark susceptibilities. In the middle panel the vertical lines are located at the same 
		temperatures of the left panel. }
\label{fig:12}
\end{figure*}

In the following we will discuss several thermodynamical quantities that allow us to 
study some observables that are accessible in lattice QCD at zero chemical potential. 
For example, full results on the QCD equation of state with 2+1 flavors at zero magnetic 
field were obtained in \cite{Borsanyi:2013bia} and, very recently, LQCD results also in 
2+1 flavors with physical quark masses and at nonzero magnetic fields were reported in 
\cite{Bali:2014kia}. By using the hadron resonance gas model the QCD equation 
of state at nonzero magnetic fields had already been performed in \cite{Endrodi:2013cs}.
Here we calculate the following: 
the pressure $P(T,B)=-[\Omega(T,B)-\Omega(0,B)]$, the energy density $ {\cal E}=TS-P$
where $S$ is the entropy density, 
the interaction measure $\Delta=({\cal E}-3P)/{T^4}$ that quantifies the deviation
from the equation of state of an ideal gas of massless constituents, 
the speed of sound squared 
\begin{equation}
v_s^2=\pc{\frac{\partial P}{\partial {\cal E}}}_V\text{,}
\end{equation}
and the specific heat 
\begin{equation}
C_V=\pc{\frac{\partial {\cal E} }{\partial T}}_V\text{.}
\end{equation}
LQCD studies show that the interaction measure remains large even at very high temperatures,
where the Stefan-Boltzmann limit is not yet reached, and thus some interaction
must still be present. 

In Fig. \ref{fig:12} we have plotted these quantities including the 
't Hooft term and a constant scalar coupling $G_s^0$ (left panel) or a
magnetic field dependent  scalar coupling $G_s(eB)$ (middle panel),
and excluding the 't Hooft term (right panel), for a magnetic field
$eB=0.3$ GeV$^2$ (the order of the maximal magnetic field
strength for the LHC \cite{Skokov:2009qp}). 

Vertical lines have been included  in all panels to indicate the position 
of the  maximum of the quark susceptibilities. In the middle panel the vertical 
lines are located at the same temperatures of the left panel to show how the 
magnetic field dependence of the coupling affects the position of the specific 
heat maximum.

As discussed before, the 't Hooft term pushes the deconfinement and
chiral transition temperatures  to larger temperatures. Moreover, for the magnetic field 
shown the $u$ and $d$ quark susceptibility maximum coincide  approximately including
the 't Hooft term but occur at quite different temperatures for $K=0$ 
(Fig. \ref{fig:12} right panel).  
This correlation between the $u$ and $d$ quarks will only be destroyed for much stronger 
magnetic fields as seen in Fig. \ref{fig:Tcs_Gd_0}.

For the three different scenarios considered it is seen that the pressure, the energy 
density and thus the interaction measure are continuous functions of the temperature 
as expected if we are in the presence of a crossover. There is a sharp increase in the 
vicinity of the transition temperature and then a tendency to saturate at the 
corresponding ideal gas limit. Excluding the 't Hooft term makes all  curves smoother.
The sharp increase occurs at lower temperatures if a magnetic field dependent coupling 
$G_s(eB)$ is considered because the transition temperatures are pushed to lower temperatures
because the interaction is weakened.  
 
The middle panels of Fig. \ref{fig:12} show the scaled specific heat $C_V/T^3$ and the 
speed of sound squared $v_s^2$ as a function of the temperature. 
The specific heat presents two peaks, caused 
by the distinct deconfinement and chiral transitions. Again, the effect of the magnetic field 
dependent scalar coupling that  pushes the peaks to lower temperatures is clearly seen. 
Moreover, there is a larger superposition between the Polyakov loop and 
$u$ and $d$ quark susceptibilities and less pronounced peaks are observed.  
The  second peak corresponding to the chiral transition is almost washed out 
when no 't Hooft term is included, due to the large superposition of the 
Polyakov loop and quark susceptibilities.

The speed of sound squared $v_s^2$ passes through a local minimum around the deconfinement 
temperature and  reaches the limit of $1/3$ (Stefan-Boltzman limit) at high temperature. 
The minimum indicates the fast change in the quark masses. 
A second inflection occurs at the chiral transition.
As expected from the previous discussion, both features are more
pronounced within the PNJL with 't Hooft term and a constant scalar
coupling. A comment that should be made is that, for the magnetic field considered $eB=0.3$ GeV$^2$,
the peak on the $s$ quark susceptibility has no effect on all the quantities represented, 
showing that the influence of the light quark sector is predominant over
the strange quark one because the restoration of the chiral symmetry already happened 
in the light quark sector.


\section{Conclusions}

In the present work we have studied the effect of the strange quark on 
the QCD phase diagram using a 2+1  PNJL model. Although under most 
conditions it is enough to consider the $u$ and $d$ quark degrees of 
freedom, both in heavy-ion collisions and neutron stars,  strangeness 
plays an important role. It is therefore important to identify the 
features of the QCD phase diagram due to the presence of strangeness
in the presence of a magnetic field.

The main property that distinguishes the $u$ and $d$ quarks from the 
$s$ quark is its mass, more than one order of magnitude larger.  
We have analyzed the effect of the current $s$ quark mass on the QCD 
phase diagram at zero chemical potential by considering several values, 
from a mass equal to that of the $u$ and $d$ quarks to a mass two times 
the $s$ quark  mass in the vacuum. Within the PNJL the 't Hooft term strongly 
mixes the flavors and, therefore, even  the properties of the $u$ and $d$ 
quarks are strongly influenced by the $s$ quark. This is easily seen 
comparing the QCD phase diagram 
features with and without the 't Hooft term when the flavors are
decoupled.

We have shown that if the mass of the $s$ quark was closer to the 
$u$ and $d$ masses its behavior in the presence of a magnetic field 
would be similar to the $d$ quark, essentially dictated by the charge. 
However, { using the current mass of the model}, the $s$ quark is much less 
sensitive to the magnetic field than the light quarks. Due to the 
't Hooft term, it has a strong influence on the light quarks at all 
temperatures for small magnetic fields and for temperatures close to 
the transition temperature and above for strong magnetic fields. 
In particular, the large mass of the $s$ quark makes the chiral 
transition of the light sector smoother and shifted to larger temperatures.

It was shown that although with a much weaker effect, the $s$ quark 
chiral transition temperature is also affected by the magnetic field 
and increases if a constant scalar coupling is used. However, if the 
magnetic dependent coupling constant proposed in \cite{Ferreira:2014kpa} 
is considered, the critical temperature associated with the $s$ quark 
decreases with $eB$. This effect, known as the inverse magnetic
catalysis, is seen on the nonmonotonic behavior of the $s$ quark 
condensate with the magnetic field.

The 't Hooft term has opposite effects when $G_s^0$ or $G_s(eB)$ are used: 
in the first case the $s$ chiral transition is almost not affected by the magnetic field 
for $0<eB<0.45$ GeV$^2$, before being washed out by the transition of the light quarks, 
while in the second case the pseudocritical temperature has a significant decrease for 
$0<eB<0.55$ GeV$^2$.
For a constant coupling the magnetic catalysis increases the $u$ and $d$ quark 
masses and the transition temperatures, bringing them close to that of the $s$ 
quark, or above for the $u$ quark and sufficiently strong fields. The flavor mixing 
induced by the 't Hooft term thus has not much effect on the $s$ quark transition. 
On the other hand, a coupling $G_s(eB)$ that gets weaker with $eB$ originates 
the IMC effect close to the transition temperature of the $u$ and $d$ quarks. 
The $s$ quark will  be strongly influenced both directly through the
weakening of $G_s$ and by the $u$ and $d$ quarks through the 
't Hooft term, so that its transition temperature will also decrease.

The identification of the $s$ quark chiral transition temperature is only 
possible below a magnetic field of the order of $eB \sim 0.45 - 0.55$ GeV$^2$, 
0.45 GeV$^2$ for a constant coupling and 0.55 GeV$^2$ for a field dependent 
coupling. For larger magnetic fields the $s$ and $d$ quark susceptibilities 
overlap too strongly.

An important effect of the large mass of the $s$ quark is to push the chiral 
transition temperatures of the $u$ and $d$ quarks to larger temperatures due 
to the mixing induced by the 't Hooft term. 
However, a magnetic field dependent scalar coupling that weakens the interaction 
for larger magnetic fields has the effect of decreasing both chiral and 
deconfinement temperatures, more the first ones than the last. In this case 
a larger overlap between the quark susceptibilities occurs and the signature 
of the crossover on thermodynamical quantities such as sound velocities or 
specific heat becomes smoother.

\vspace{1cm}
{\bf ACKNOWLEGMENTS}: 
This work was partially supported by Projects PTDC/FIS/113292/2009, PEst-OE/FIS/UI0405/2014 and
CERN/FP/123620/2011 developed under the initiative QREN financed by the UE/FEDER 
through the program COMPETE $-$ ``Programa Operacional Factores de Competitividade'' 
and by Grant No. SFRH/BD/51717/2011.


\end{document}